\documentclass[preprint]{elsarticle}

\usepackage{lineno,hyperref}
\modulolinenumbers[5]

\usepackage{amsmath} 
\usepackage{amssymb} 

\journal{J. Magn. Magn. Mater.}

\begin{document}

\begin{frontmatter}

\title{Magnetic excitations and transport properties in frustrated ferromagnetic chain}

\author{Hiroaki Onishi}
\address{Advanced Science Research Center, Japan Atomic Energy Agency, Tokai, Ibaraki 319-1195, Japan}

\begin{abstract}
To clarify the transport property mediated by bound magnons in a spin nematic state,
we investigate the spin Drude weight of
a spin-1/2 $J_{1}$-$J_{2}$ Heisenberg chain
with ferromagnetic $J_{1}$ and antiferromagnetic $J_{2}$ in a magnetic field
by numerical diagonalization.
At zero magnetic field,
numerical results with up to 24 sites suggest that
the spin Drude weight would be zero in the thermodynamic limit,
indicating diffusive spin transport.
With increasing the magnetic field,
the spin Drude weight is enhanced at low temperatures,
indicating that bound magnons would contribute to the spin transport.
\end{abstract}

\begin{keyword}
frustrated ferromagnetic chain \sep
spin nematic state \sep
bound magnons \sep
spin transport
\end{keyword}

\end{frontmatter}


\section{Introduction}

Transport properties of low-dimensional magnetic materials have attracted great interest
not only for studying fundamental problems
but also for seeking possible applications to spintronics
\cite{Zotos2005,Baeriswyl-book2004,Zutic2004,Maekawa-book2017}.
In ordinary magnetic insulators,
elementary excitation from a magnetic ordered state is described by a magnon,
which is a quantized spin wave.
Magnons carry spin current and thermal current.
In contrast,
frustrated quantum magnets frequently exhibit non-trivial ground states and elementary excitations,
since the combined effects of frustration and quantum fluctuations disturb magnetic order.
Thus we expect that novel types of carriers would contribute to spin and thermal transport phenomena.

As a prominent example of exotic ground states and excitations caused by frustration,
we focus on a spin nematic state in a spin-1/2 $J_{1}$-$J_{2}$ Heisenberg chain
with ferromagnetic $J_{1}$ and antiferromagnetic $J_{2}$
in a magnetic field.
At high fields,
the ground state is spin nematic with quasi-long-range quadrupole correlations,
while it is also characterized by the formation of bound magnons
\cite{Chubukov1991,Kecke2007,Vekua2007,Hikihara2008,Sudan2009}.
Since the quadrupole and longitudinal spin correlations are both quasi-long-ranged,
we refer to this phase as the spin-nematic/spin-density-wave (SN/SDW) phase.
Many theoretical efforts have been made to clarify dynamical properties
\cite{Sato2009,Sato2011,Onishi2015a,Onishi2015b,Smerald2015,Furuya2017,Furuya2018,Onishi2018,Ramos2018},
whereas the relevance of magnetic excitations governed by bound magnons
to transport properties has not been understood yet.

Regarding magnetic excitations,
gapless longitudinal and gapped transverse spin excitation spectra occur,
in accordance with quasi-long-range longitudinal and short-range transverse spin correlations, respectively,
while we find gapless quadrupole excitation
due to quasi-long-range quadrupole correlations.
Here, it is interesting to note that
bound magnons can be created without energy loss
due to the gapless quadrupole excitation,
so that bound magnons would carry spin current and thermal current.

Theoretically,
the Drude weight,
which is the zero-frequency component of the conductivity,
is an important quantity to clarify whether the transport is ballistic or diffusive.
For an integrable XXZ Heisenberg chain,
it is now well established that
the thermal transport is ballistic via analytical and numerical approaches
\cite{Zotos1997,Klumper2002,Sakai2003,Heidrich-Meisner2003}.
The spin transport is ballistic in the gapless XY phase,
rigorously proved by using the Mazur inequality
\cite{Prosen2011,Prosen2013},
while diffusive in the gapped regime
\cite{Znidaric2011,Karrasch2012}.
The relevance of the integrability to the ballistic transport has been studied
\cite{Heidrich-Meisner2003,Jung2007,Sirker2011,Wu2011,Znidaric2013,Karrasch2015},
while it is still an interesting open issue to understand transport properties
of non-integrable frustrated spin systems

In this paper,
to gain an insight into the spin transport property
in the spin nematic state of the frustrated ferromagnetic chain,
we investigate the spin Drude weight numerically.
We argue that bound magnons would contribute to the spin transport,
based on detailed analyses of the temperature and magnetic field dependencies of the spin Drude weight.

\section{Model and Method}

We consider a spin-1/2 $J_{1}$-$J_{2}$ Heisenberg chain of $N$ sites,
described by
\begin{equation}
  H =
  J_{1} \sum_{i} \mbox{\boldmath $S$}_{i} \cdot \mbox{\boldmath $S$}_{i+1}
  + J_{2} \sum_{i} \mbox{\boldmath $S$}_{i} \cdot \mbox{\boldmath $S$}_{i+2}
  - h \sum_{i} S_{i}^{z},
\label{eq: H}
\end{equation}
where $J_{1}(<0)$ denotes the ferromagnetic nearest-neighbor exchange interaction,
$J_{2}(>0)$ the antiferromagnetic next-nearest-neighbor exchange interaction,
and $h$ the magnetic field.
Throughout the paper, we set $J_{2}=1$ and take it as the energy unit.
We adopt the periodic boundary condition.

The spin conductivity $\sigma$ is defined as the linear response of
the spin current
to the magnetic field gradient.
Within the linear response theory,
the spin conductivity is obtained from spin current correlation functions
in thermal equilibrium without the magnetic field gradient.
The real part of frequency dependent $\sigma(\omega)$ is expressed by
a $\delta$ function at $\omega=0$
and a regular part at finite frequency $\sigma_{\mathrm{reg}}(\omega)$
as
$\mathrm{Re} \, \sigma(\omega)=D_{\mathrm{s}}\delta(\omega)+\sigma_{\mathrm{reg}}(\omega)$,
where $D_{\mathrm{s}}$ is the spin Drude weight.
Based on the Kubo formula,
the spin Drude weight reads
\begin{equation}
  D_{\mathrm{s}} =
  \frac{\pi\beta}{N}
  \sum_{m,n (E_{m}=E_{n})}
  \frac{\exp(-\beta E_{n})}{Z}
  \vert \langle m \vert j_{\mathrm{s}} \vert n \rangle \vert^{2},
\label{eq: Ds}
\end{equation}
where
$j_{\mathrm{s}}$ is the spin current operator,
$\vert n \rangle$ is the eigenvector with the eigenenergy $E_{n}$,
$\beta=1/k_{\mathrm{B}}T$
with $k_{\mathrm{B}}$ and $T$ being the Boltzmann factor and the temperature, respectively,
and
$Z=\sum_{n}\exp(-\beta E_{n})$.
The local spin current operator $j_{\mathrm{s},i}$ satisfies
the continuity equation
$\partial_{t}S_{i}^{z}+j_{\mathrm{s},i+1}-j_{\mathrm{s},i}=0$
and the equation of motion
$\partial_{t} S_{i}^{z}=\mathrm{i} [H,S_{i}^{z}]$,
leading to
$j_{\mathrm{s},i}=\mathrm{i}[H_{i-1}+H_{i-2},S_{i}^{z}+S_{i+1}^{z}]$
with
$H_{i}=J_{1}\mbox{\boldmath $S$}_{i}\cdot\mbox{\boldmath $S$}_{i+1}+J_{2}\mbox{\boldmath $S$}_{i}\cdot\mbox{\boldmath $S$}_{i+2}-hS_{i}^{z}$.
Thus the total spin current $j_{\mathrm{s}}=\sum_{i}j_{\mathrm{s},i}$ is given by
$j_{\mathrm{s}}=J_{1}\sum_{i}(\mbox{\boldmath $S$}_{i}\times\mbox{\boldmath $S$}_{i+1})^{z}+2J_{2}\sum_{i}(\mbox{\boldmath $S$}_{i}\times\mbox{\boldmath $S$}_{i+2})^{z}$.

We calculate the spin Drude weight by the numerical diagonalization method.
We obtain all eigenenergies and eigenvectors,
and evaluate the thermal average in Eq.~(\ref{eq: Ds}).
The Hamiltonian is block-diagonalized
by using the conservation of the total magnetization and the translational symmetry,
so that the maximum dimension is 9~252 for 20 sites,
32~066 for 22 sites,
and
112~720 for 24 sites.
In the present study,
we examine the temperature and magnetic field dependencies of $D_{\mathrm{s}}$
at $(J_{1},J_{2})=(-1,1)$
for periodic chains of up to $N=24$.
Note that the ground state at $(J_{1},J_{2})=(-1,1)$ varies with the magnetic field
\cite{Hikihara2008,Sudan2009,Furukawa2012}:
a gapped dimer state at zero magnetic field,
a vector chiral state in a narrow range of low magnetic fields,
a SN/SDW state up to a saturation field $h_{\mathrm{sat}}^{T=0}\simeq 1.25$,
and a forced ferromagnetic state above the saturation field.

\section{Results}

\begin{figure}[t]
\begin{center}
\includegraphics[scale=0.7]{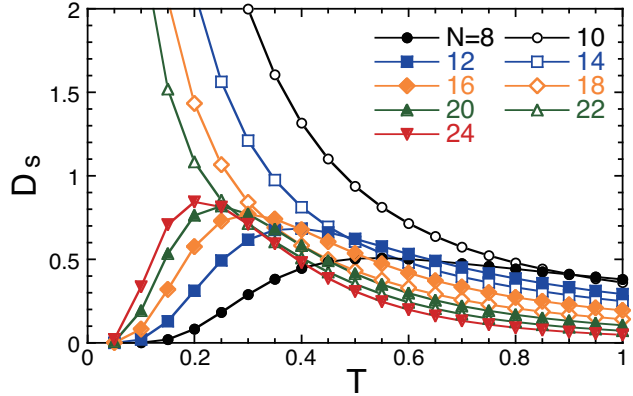}
\end{center}
\caption{
The temperature dependence of the spin Drude weight $D_{\mathrm{s}}$
at $J_{1}=-1$, $J_{2}=1$, and $h=0$.
Closed symbols denote data for $N=4n$ with integer $n$,
while open symbols are for $N=4n+2$.
}
\label{fig1}
\end{figure}

In Fig.~1,
we show the temperature dependence of $D_{\mathrm{s}}$ at zero magnetic field,
where the ground state is gapped,
for various system sizes up to $N=24$.
At a glance,
data for $N=4n$ with integer $n$ (closed symbols)
and those for $N=4n+2$ (open symbols) exhibit different behavior.
This is because the system is regarded as two antiferromagnetic $J_{2}$ chains
connected by ferromagnetic $J_{1}$,
and each antiferromagnetic chain includes
even sites for $N=4n$ and odd sites for $N=4n+2$.
Note that the finite-size correction is large
even when we increase the system size up to $N=24$,
so that it is difficult to reach a definite conclusion
on the behavior in the thermodynamic limit.

For $N=4n$,
we see that $D_{\mathrm{s}}$ has a broad peak
and decays to zero as the temperature approaches zero
due to the excitation gap in the finite-size system.
The peak is shifted toward low temperature as the system size increases.
We also find that
$D_{\mathrm{s}}$ decreases with $N$ at high temperatures,
while it changes to increase with $N$ at low temperatures.
There is a crossover temperature,
where the change between increase and decrease of $D_{\mathrm{s}}$ with $N$ occurs.
For instance,
looking at data of $N=8$ and $N=12$,
we observe that 
$D_{\mathrm{s}}(N=8)>D_{\mathrm{s}}(N=12)$ for $T>0.675$,
while $D_{\mathrm{s}}(N=8)<D_{\mathrm{s}}(N=12)$ for $T<0.675$,
where the crossover temperature is at $T=0.675$.
If the crossover temperature remains at finite temperature in the thermodynamic limit,
$D_{\mathrm{s}}$ should be finite below the crossover temperature.
However,
the crossover temperature shifts toward low temperature
and seems to approach zero temperature with $N$,
suggesting that $D_{\mathrm{s}}$ would become zero in the thermodynamic limit,
i.e., diffusive spin transport.
Note that these observations are quite similar to those
in the gapped regime of the antiferromagnetic $J_{1}$-$J_{2}$ chain
\cite{Heidrich-Meisner2003}.

In contrast, for $N=4n+2$,
$D_{\mathrm{s}}$ grows as the temperature decreases.
Since the two antiferromagnetic $J_{2}$ chains include odd sites,
we assume that an effective spin-1/2 is formed in each of the two chains
and they form an effective spin-1 via the ferromagnetic $J_{1}$ coupling.
Such an effective spin-1 would sensitively respond to the magnetic field gradient,
leading to the enhancement of $D_{\mathrm{s}}$.
We also find that
$D_{\mathrm{s}}$ decreases with $N$ in the whole temperature range analyzed.
This is suggestive of vanishing $D_{\mathrm{s}}$ in the thermodynamic limit,
consistent with data for $N=4n$.

\begin{figure}[t]
\begin{center}
\includegraphics[scale=0.7]{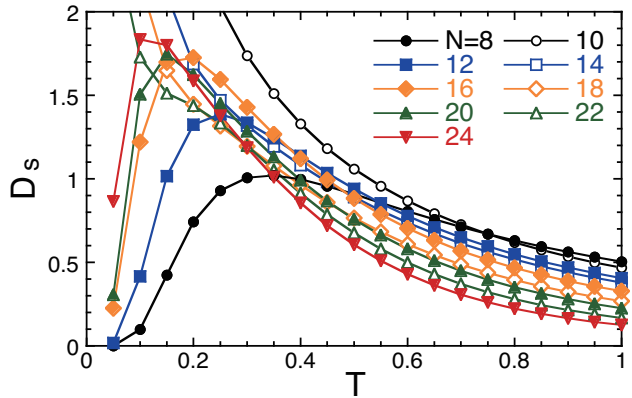}
\end{center}
\caption{
The temperature dependence of the spin Drude weight $D_{\mathrm{s}}$
at $J_{1}=-1$, $J_{2}=1$, and $h=1$,
where the ground state is SN/SDW.
Closed symbols denote data for $N=4n$ with integer $n$,
while open symbols are for $N=4n+2$.
}
\label{fig2}
\end{figure}

\begin{figure}[t]
\begin{center}
\includegraphics[scale=0.7]{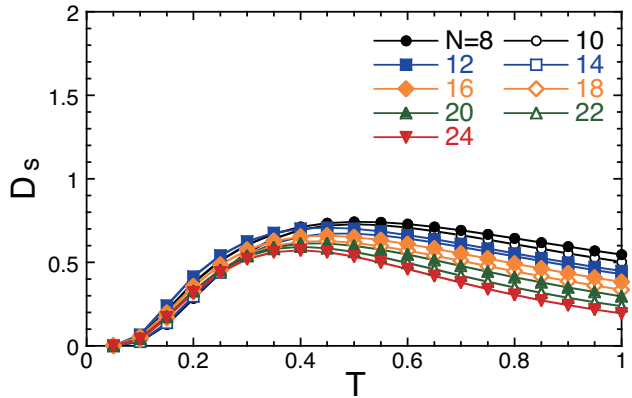}
\end{center}
\caption{
The temperature dependence of the spin Drude weight $D_{\mathrm{s}}$
at $J_{1}=-1$, $J_{2}=1$, and $h=1.5$,
where the ground state is ferromagnetic.
Closed symbols denote data for $N=4n$ with integer $n$,
while open symbols are for $N=4n+2$.
}
\label{fig3}
\end{figure}

Now we move on to the investigation at finite magnetic field.
In Fig.~2,
we show the temperature dependence of $D_{\mathrm{s}}$ at $h=1$,
where the ground state is SN/SDW.
Note that the difference between $N=4n$ and $N=4n+2$ can be seen
in the same way as the case of zero magnetic field.
Here,
we find that $D_{\mathrm{s}}$ is enhanced particularly at low temperatures
as compared with the case of zero magnetic field.
Figure~3 presents $D_{\mathrm{s}}$ at $h=1.5$,
where the ground state is ferromagnetic.
We see that $D_{\mathrm{s}}$ decays to zero as the temperature decreases
regardless of whether the system size is $N=4n$ or $N=4n+2$
because of the ferromagnetic gap.

\begin{figure}[t]
\begin{center}
\includegraphics[scale=0.7]{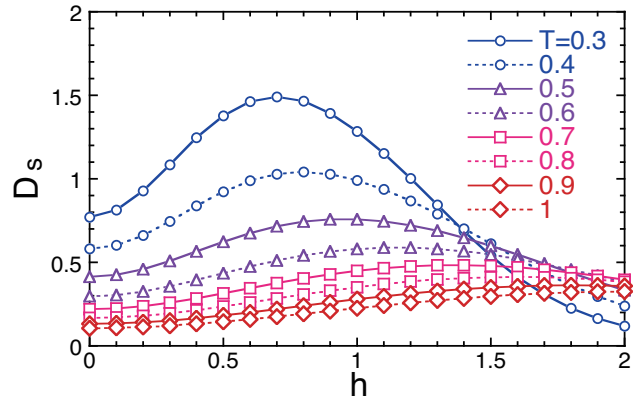}
\end{center}
\caption{
The magnetic field dependence of the spin Drude weight $D_{\mathrm{s}}$
at $J_{1}=-1$, $J_{2}=1$, and several temperatures.
The system size is $N=20$.
}
\label{fig4}
\end{figure}

Let us look into the magnetic field dependence in more detail.
In Fig.~4,
we plot $D_{\mathrm{s}}$ as a function of the magnetic field
at several temperatures.
At low temperatures, e.g., $T=0.3$,
$D_{\mathrm{s}}$ is strongly enhanced with increasing the magnetic field,
while it is suppressed after showing a peak at around $h=0.7$.
As the temperature increases,
the peak structure is broadened and the spectral weight is distributed to higher magnetic field.
We mention that
the spin current correlation function should be zero
above the saturation field $h_{\mathrm{sat}}^{T=0}\simeq 1.25$ at zero temperature,
since the multiplication of the spin current operator to the ferromagnetic ground state
yields zero.
At finite temperature,
$D_{\mathrm{s}}$ would become finite above $h_{\mathrm{sat}}^{T=0}$
due to thermal fluctuations.
Note here that the SN/SDW state is the ground state
and it should have contribution to observables at low temperature.
In this sense,
the enhancement of $D_{\mathrm{s}}$ at low temperature
would be a precursory phenomenon of the SN/SDW state at zero temperature,
suggesting that bound magnons would contribute to the spin transport.

\section{Summary}

We have studied the spin transport property of
the spin-1/2 $J_{1}$-$J_{2}$ Heisenberg chain
with ferromagnetic $J_{1}$ and antiferromagnetic $J_{2}$
in the magnetic field
by the numerical diagonalization method.
At zero magnetic field,
the spin Drude weight seems to become zero in the thermodynamic limit,
suggesting diffusive spin transport.
In a magnetic field range where the ground state is spin nematic,
the spin Drude weight is enhanced at low temperatures,
indicating that bound magnons would contribute to the spin transport.

\section*{Acknowledgments}

This work was supported by JSPS KAKENHI Grant Number JP16K05494.
Part of computations were carried out on the supercomputers at
the Japan Atomic Energy Agency and
the Institute for Solid State Physics, the University of Tokyo.

\section*{References}


\begin{thebibliography}{}

\bibitem{Zotos2005}
X. Zotos,
J. Phys. Soc. Jpn. \textbf{74} Suppl., 173 (2005).

\bibitem{Baeriswyl-book2004}
\textit{Strong Interactions in Low Dimensions},
ed. D. Baeriswyl and L. Degiorgi
(Kluwer Academic Publishers, Dordrecht, 2004).

\bibitem{Zutic2004}
I. \v{Z}uti\'{c}, J. Fabian, and S. D. Sarma,
Rev. Mod. Phys. \textbf{76}, 323 (2004).

\bibitem{Maekawa-book2017}
\textit{Spin Current},
ed. S. Maekawa, S. O. Valenzuela, E. Saitoh, and T. Kimura
(Oxford University Press, Oxford, 2017).

\bibitem{Chubukov1991}
A. V. Chubukov,
Phys. Rev. B \textbf{44}, 4693 (1991).

\bibitem{Kecke2007}
L. Kecke, T. Momoi, and A. Furusaki,
Phys. Rev. B \textbf{76}, 060407(R) (2007).

\bibitem{Vekua2007}
T. Vekua, A. Honecker, H. J. Mikeska, and F. Heidrich-Meisner,
Phys. Rev. B \textbf{76}, 174420 (2007).

\bibitem{Hikihara2008}
T. Hikihara, L. Kecke, T. Momoi, and A. Furusaki,
Phys. Rev. B \textbf{78}, 144404 (2008).

\bibitem{Sudan2009}
J. Sudan, A. L\"uscher, and A. M. L\"auchli,
Phys. Rev. B \textbf{80}, 140402(R) (2009).

\bibitem{Sato2009}
M. Sato, T. Momoi, and A. Furusaki,
Phys. Rev. B \textbf{79}, 060406(R) (2009).

\bibitem{Sato2011}
M. Sato, T. Hikihara, and T. Momoi,
Phys. Rev. B \textbf{83}, 064405 (2011).

\bibitem{Onishi2015a}
H. Onishi,
J. Phys. Soc. Jpn. \textbf{84}, 083702 (2015).

\bibitem{Onishi2015b}
H. Onishi,
J. Phys.: Conf. Ser. \textbf{592}, 012109 (2015).

\bibitem{Smerald2015}
A. Smerald, H. T. Ueda, and N. Shannon,
Phys. Rev. B \textbf{91}, 174402 (2015).

\bibitem{Furuya2017}
S. C. Furuya,
Phys. Rev. B \textbf{95}, 014416 (2017).

\bibitem{Furuya2018}
S. C. Furuya and T. Momoi,
Phys. Rev. B \textbf{97}, 104411 (2018).

\bibitem{Onishi2018}
H. Onishi,
Physica B \textbf{536}, 346 (2018).

\bibitem{Ramos2018}
F. B. Ramos, S. Eli\"{e}ns, and R. G. Pereira,
arXiv:1805.10612.

\bibitem{Zotos1997}
X. Zotos, F. Naef, and P. Prelovsek,
Phys. Rev. B \textbf{55}, 11029 (1997).

\bibitem{Klumper2002}
A. Klumper and K. Sakai,
J. Phys. A: Math. Gen. \textbf{35}, 2173 (2002).

\bibitem{Sakai2003}
K. Sakai and A. Klumper,
J. Phys. A: Math. Gen. \textbf{36}, 11617 (2003).

\bibitem{Heidrich-Meisner2003}
F. Heidrich-Meisner, A. Honecker, D. C. Cabra, and W. Brenig,
Phys. Rev. B \textbf{68}, 134436 (2003).

\bibitem{Prosen2011}
T. Prosen,
Phys. Rev. Lett. \textbf{106}, 217206 (2011).

\bibitem{Prosen2013}
T. Prosen and E. Ilievski,
Phys. Rev. Lett. \textbf{111}, 057203 (2013).

\bibitem{Znidaric2011}
M. \v{Z}nidari\v{c},
Phys. Rev. Lett. \textbf{106}, 220601 (2011).

\bibitem{Karrasch2012}
C. Karrasch, J. H. Bardarson, and J. E. Moore,
Phys. Rev. Lett. \textbf{108}, 227206 (2012).

\bibitem{Jung2007}
P. Jung and A. Rosch,
Phys. Rev. B \textbf{76}, 245108 (2007).

\bibitem{Sirker2011}
J. Sirker, R. G. Pereira, and I. Affleck,
Phys. Rev. B \textbf{83}, 035115 (2011).

\bibitem{Wu2011}
J. Wu and M. Berciu,
Phys. Rev. B \textbf{83}, 214416 (2011).

\bibitem{Znidaric2013}
M. \v{Z}nidari\v{c},
Phys. Rev. Lett. \textbf{110}, 070602 (2013).

\bibitem{Karrasch2015}
C. Karrasch, D. M. Kennes, and F. Heidrich-Meisner,
Phys. Rev. B \textbf{91}, 115130 (2015).

\bibitem{Furukawa2012}
S. Furukawa, M. Sato, S. Onoda, and A. Furusaki,
Phys. Rev. B \textbf{86}, 094417 (2012).

\end{thebibliography}

\end{document}